\documentclass[preprint,showpacs,superscriptaddress,preprintnumbers,amsmath,amssymb]{revtex4}

\usepackage{graphicx}
\usepackage{color}

\begin{document}

\title{Lieb-Liniger gas in a constant-force potential}

\author{D. Juki\'c}
\email{djukic@phy.hr}
\affiliation{Department of Physics, University of Zagreb, Bijeni\v cka c. 32, 10000 Zagreb, Croatia}
\author{S. Gali\'c}
\affiliation{Department of Physics, University of Zagreb, Bijeni\v cka c. 32, 10000 Zagreb, Croatia}
\author{R. Pezer}
\affiliation{Faculty of Metallurgy, University of Zagreb, Aleja narodnih heroja 3, 44103 Sisak, Croatia}
\author{H. Buljan}
\affiliation{Department of Physics, University of Zagreb, Bijeni\v cka c. 32, 10000 Zagreb, Croatia}

\date{\today}

\begin{abstract}
We use Gaudin's Fermi-Bose mapping operator to calculate exact solutions for 
the Lieb-Liniger model in a linear (constant-force) potential (the constructed 
exact stationary solutions are referred to as the Lieb-Liniger-Airy wave functions). 
The ground-state properties of the gas in the wedgelike trapping potential 
are calculated in the strongly interacting regime 
by using Girardeau's Fermi-Bose mapping and the pseudopotential approach in the 
$1/c$ approximation ($c$ denotes the strength of the interaction). 
We point out that quantum dynamics of Lieb-Liniger wave packets in the 
linear potential can be calculated by employing an $N$-dimensional 
Fourier transform as in the case of free expansion. 
\end{abstract}

\pacs{03.75.Kk, 05.30.-d, 67.85.De}
\maketitle

\section{Introduction}
\label{sec:intro}

One-dimensional (1D) Bose gases hold great potential for studying the physics of 
interacting quantum many-body systems.
In recent years, a growing interest in studies of theoretical 1D models, first 
introduced by Lieb and Liniger \cite{Lieb1963} and Girardeau \cite{Girardeau1960}, 
has largely been inspired by experimental progress in realizing 
these models with ultracold atomic gases \cite{OneD,TG2004,Kinoshita2006,Hofferberth2007,
Amerongen2008}. In experiments, an effectively one-dimensional system is achieved in 
elongated and transversely tight atomic wave guides, loaded with ultracold atoms, 
where transverse excitations are strongly suppressed \cite{OneD,TG2004,Kinoshita2006,
Hofferberth2007,Amerongen2008}.
These atomic gases are well described by the Lieb-Liniger model \cite{Lieb1963} - 
a system of identical Bose particles in 1D which interact via 
$\delta$-function interactions of strength $c$.
In the limit of infinite interaction strength ($c \rightarrow \infty$), 
the Bose particles can be described by the Tonks-Girardeau model \cite{Girardeau1960}, 
describing an "impenetrable" Bose gas. 
This regime occurs when effective interactions are strong, whereas temperatures and 
linear densities are low \cite{Olshanii,Petrov,Dunjko}.
In passing, we note that the Lieb-Liniger model can also be realized in the field of
quantum optics \cite{Chang2008}.
An ubiquitous part of the ultracold atomic experiments is the gravitational force, which 
can be nullified in a precise experiment, where 1D atomic wave guides are 
horizontal. However, to the best of our knowledge, 
exact solutions for the Lieb-Liniger gas (consisting of an arbitrary number of particles) in the presence of a constant external force have not been constructed yet. 

An important breakthrough in the context of interacting Bose gases 
was achieved in 1963 when Lieb and Liniger found exact eigenstates 
for a 1D Bose gas with $\delta$-function (repulsive) interactions of arbitrary 
strength $c$  \cite{Lieb1963}. The underlying structure of the eigenstates reveals the Bethe 
ansatz, and each eigenstate is determined by a set of $N$ quasimomenta, where $N$ 
denotes the number of particles. 
If the quasimomenta obey a particular set of transcendental equations \cite{Lieb1963}, 
the wave functions will obey periodic boundary conditions. 
This breakthrough was followed by studies of the model with attractive 
interactions \cite{McGuire1964}. A superposition of Lieb-Liniger eigenstates,
where again quasimomenta obey a particular set of transcendental equations, 
can be used to construct the ground state in an infinitely deep box \cite{Gaudin1971}. 
More recent studies of the Lieb-Liniger model (e.g., \cite{Muga1998,
Sakmann2005,Batchelor2005,Forrester2006,Kinezi2006,Sykes2007,Kanamoto2009,Gritsev2010}) 
are mainly motivated by the growing experimental progress 
\cite{OneD,TG2004,Kinoshita2006,Hofferberth2007,Amerongen2008}. 
It should be noted that all of the aforementioned studies focus on a system 
where an external potential is zero in a given region of the $x$ space, and 
boundary conditions are imposed on the wave function at the border(s) of this 
region (e.g., periodic \cite{Lieb1963}, semi-infinite line \cite{Gaudin1971}, and
infinitely deep box \cite{Gaudin1971}).
For other external potentials, there are studies of few-body systems \cite{Sen1988,Busch1998}. 
Exact solutions were presented for $N=2$ particles in a harmonic trap in Ref. 
\cite{Busch1998}; solutions for $N=2$ and $N=3$ particles in a linear potential 
were presented in \cite{Sen1988}.

One approach in attempting to broaden the set of known exact solutions is by 
using Gaudin's Fermi-Bose mapping operator (call it $\hat O_c$) \cite{Gaudin1983}, 
which is applicable also to find time-dependent Lieb-Liniger wave functions 
\cite{Gaudin1983,Buljan2008,Jukic2008,Jukic2009,Pezer2009}. 
In the limit of the Tonks-Girardeau gas $c\rightarrow \infty$, the Fermi-Bose mapping 
(which was discovered in 1960 \cite{Girardeau1960}) can be utilized in any external 
potential \cite{Girardeau1960} and for time-dependent problems \cite{Girardeau2000} 
(for a review, see also \cite{Yukalov2005}).
For the finite coupling Lieb-Liniger gas ($c$ finite), the method of Gaudin has 
been shown to be valid in the absence of any external potential (i.e., on an infinite 
line \cite{Buljan2008}), and has been used to study free expansion from localized initial
conditions \cite{Jukic2008,Jukic2009}; in this case the time-dependent wave function 
can be calculated via an $N$-dimensional Fourier transform. Interestingly, 
such a transform can be also utilized for a Lieb-Liniger gas reflecting from the 
wall \cite{Jukic2010}. 
However, Gaudin's method (at least in its current form) is not applicable to find 
eigenstates of a Lieb-Liniger gas in generic trapping potentials $V(x)$ (such as 
the harmonic oscillator); technically, this arises because the differential operator 
$\hat O_c$ does not generally commute with such potentials. 

Here, we study the Lieb-Liniger model in the constant-force (linear) potential. 
Exact stationary solutions for this system are constructed (we call these wave 
functions the Lieb-Liniger-Airy states) by employing Gaudin's operator $\hat O_c$. 
The construction is enabled by the fact that this operator commutes with the linear 
(constant-force) potential.
We calculate the ground-state properties of the Lieb-Liniger gas in the wedgelike 
potential [$V(x)=\alpha x$ for $x>0$ ($\alpha>0$), and $\infty$ otherwise] in the 
strongly interacting regime. This is achieved in the Tonks-Girardeau regime 
and below that regime in $1/c$ approximation by employing the pseudopotential 
approach \cite{Sen99-03}. Finally, we point out that the time-dependent Lieb-Liniger
wave packets in the linear potential can be calculated via an $N$-dimensional Fourier transform.

\section{Lieb-Liniger-Airy states}
\label{sec:LLmodel}
The Lieb-Liniger model describes $N$ identical bosons in one spatial dimension which 
interact via (delta-function) contact potential \cite{Lieb1963}. 
In this section we consider this system placed in a linear external potential. 
The stationary Schr\"odinger equation for the many-body wave function 
$\psi_B(x_1,\ldots,x_N)$ in such a system is 
\begin{equation}
E \psi_B =
- \sum_{i=1}^{N}\frac{\partial^2 \psi_B}{\partial x_i^2} 
+ \sum_{1\leq i < j \leq N} 2c\,\delta(x_i-x_j)\psi_B 
+ \alpha \sum_{i=1}^N x_i  \psi_B,
\label{LLmodel}
\end{equation}
where $c>0$ denotes the strength of the interaction, and $\alpha>0$ is the constant 
external force. Solutions of Eq.  (\ref{LLmodel}) for a single particle $(N=1)$ 
are the Airy functions. For this reason, in what follows, we will call the solutions of 
Eq. (\ref{LLmodel}) for $N>1$ the Lieb-Liniger-Airy (LLA) states (we are interested only 
in those solutions which decay to zero when $x\rightarrow \infty$). 
The constant force in ultracold atomic experiments can arise from the gravity force 
(e.g., if the one-dimensional atomic wave guides are tilted with respect to gravity). 

In what follows, we will demonstrate that LLA states can be constructed 
via Gaudin's Fermi-Bose transformation \cite{Gaudin1983}. 
Because of the bosonic symmetry of the wave functions, one can consider 
only the fundamental permutation sector of the coordinate space 
$R_1:x_1<x_2<\ldots<x_N$. Within this sector, the Schr\" odinger equation 
(\ref{LLmodel}) reads
\begin{equation}
E \psi_B=
- \sum_{i=1}^{N}\frac{\partial^2 \psi_B}{\partial x_i^2} 
+ \alpha \sum_{i=1}^N x_i  \psi_B.
\label{LLmodel2}
\end{equation}
The interaction term is taken into account as a boundary condition (the so 
called cusp condition), which is imposed upon $\psi_B$ at the borders of $R_1$
(i.e., when two particles touch \cite{Lieb1963}):
\begin{equation}
\left [
1-\frac{1}{c}
\left (
\frac{\partial}{\partial x_{j+1}}-\frac{\partial}{\partial x_j}
\right)
\right]_{x_{j+1}=x_j}\psi_B=0.
\label{interactions}
\end{equation}
Equation (\ref{LLmodel2}) holds in all other permutation sectors, whereas 
the interaction cusp (\ref{interactions}) can be re-expressed on the borders of 
other sectors as well.
To construct the LLA states we utilize Gaudin's Fermi-Bose mapping operator \cite{Gaudin1983},
\begin{equation}
\hat O_c=\prod_{1\leq i < j \leq N}
\left[
\mbox{sgn}(x_j-x_i)+\frac{1}{c}
\left(
\frac{\partial}{\partial x_{j}}-
\frac{\partial}{\partial x_{i}}
\right)
\right],
\label{oO}
\end{equation}
which acts upon an antisymmetric (fermionic) wave function $\psi_F$. 
The wave function $\psi_F$ must obey the Schr\" odinger equation for 
noninteracting spinless fermions in the linear potential:
\begin{equation}
E \psi_F =
- \sum_{i=1}^{N}\frac{\partial^2 \psi_F}{\partial x_i^2} 
+ \alpha \sum_{i=1}^N x_i  \psi_F.
\label{fermionic}
\end{equation}
The wave function $\psi_F$ can be written in the form of Slater 
determinant with Airy functions as entries:
\begin{equation} 
\psi_F = \alpha^{-\frac{N}{6}} \frac{1}{\sqrt{N!}} \det_{i,j=1}^{N}
\mbox{Ai}(\alpha^{\frac{1}{3}} x_j - \alpha^{-\frac{2}{3}} E_i),
\end{equation}
where $E=\sum_{i=1}^{N} E_i$. 

The LLA states [i.e., solutions of the Schr\"odinger Eq. (\ref{LLmodel2}), together with 
the cusp condition (\ref{interactions})], are given by 
\begin{equation}
\psi_{B,c}= {\mathcal N}_{c} \hat O_c \psi_F,
\label{ansatz}
\end{equation}
where ${\mathcal N}_{c}$ is the normalization constant.
It is known that all wave functions of the form (\ref{ansatz}) obey the cusp conditions
throughout the configuration space \cite{Gaudin1983,Buljan2008,Jukic2008}.
To show that $\psi_{B,c}$ is also a solution of Eq. (\ref{LLmodel2}), it is sufficient 
to prove that the following commutators are zero: 
$\left[ \sum_i \partial^2/\partial x_i^2,\hat O_c \right]=0$ 
and $\left[ \sum_i x_i,\hat O_c \right]=0$; this is sufficient because $\psi_F$ obeys Eq. 
(\ref{fermionic}).
The first commutator is trivially
satisfied, and therefore we are left to verify that 
\begin{equation}
\left[ \sum_i x_i,\hat O_c \right]=0.
\label{secondcom}
\end{equation}
As a first step, we restrict ourselves to the case of two particles, $N=2$. By using 
$\left[x_j,  \partial/\partial x_i  \right]=-\delta_{j,i}$,
we have
\begin{equation} 
\left[ x_1+x_2,\mbox{sgn}(x_2-x_1)+
  \frac{1}{c}\left(\frac{\partial}{\partial x_2}-\frac{\partial}{\partial x_1} 
             \right)
\right] =
  \frac{1}{c} \left[ x_2,\frac{\partial}{\partial x_2} 
              \right]-
  \frac{1}{c} \left[ x_1,\frac{\partial}{\partial x_1} \right] = 0.
\end{equation}
Now we generalize this for any number of particles $N$.
Let us write the differential operator as
$\hat O_c=\prod_{1\leq i < j \leq N} \hat B_{i,j}$,
where
\begin{equation}
\hat B_{i,j}=\left[
\mbox{sgn}(x_j-x_i)+\frac{1}{c}
\left(
\frac{\partial}{\partial x_{j}}-
\frac{\partial}{\partial x_{i}}
\right)
\right]
\label{Bij}.
\end{equation}
A general expression,
$\left[ \hat V, \prod_{l=1}^M \hat W_l \right]=
\sum_{l=1}^M  \hat W_1 \cdots \hat W_{l-1} \left[ \hat V,\hat W_l \right]
\hat W_{l+1} \cdots \hat W_M,
$
valid for operators $\hat V$ and $\hat W_l$, $l=1,\ldots,M$,
enables us to write the required commutator for the case of $N$ particles: 
\begin{equation} 
\left[ \sum_k x_k,\hat O_c \right] = 
\sum_{i<j}  \hat B_{N-1,N} \cdots \left[ \sum_k x_k,\hat B_{i,j} \right]
\cdots \hat B_{1,2}.
\end{equation}
Now Eq. (\ref{secondcom}) follows immediately because for any 
$\hat B_{i,j}$ we have $\left[ \sum_k x_k,\hat B_{i,j} \right]=
\left[x_i+x_j,\hat B_{i,j} \right]=0$, as is verified for the $N=2$ case.
This completes the proof that the wave function $\psi_{B,c}$ defined 
in (\ref{ansatz}) is a solution of Eq. (\ref{LLmodel}).

In this section we have found exact closed form solutions of Eq. \eqref{LLmodel}.
We point out that the eigenstates \eqref{ansatz} with total energy $E$
are degenerate, because the choice of single particle energies $E_i$ for which
$E=\sum_{i=1}^{N} E_i$ is not unique. 
By superposition of degenerate eigenstates \eqref{ansatz}, one can construct 
eigenstates which are of different mathematical form. 
In \cite{Sen1988} the authors study Eq. \eqref{LLmodel} for $N=2$ and $N=3$ particles.
They constructed solutions by introducing a new set of coordinates and separating
Eq. \eqref{LLmodel}. For $N=2$ they separate the center of mass and relative
motion. Their solution for a given energy can be written as a superposition of 
eigenstates \eqref{ansatz}. For $N=3$ the procedure in \cite{Sen1988} becomes more 
cumbersome, which clearly points out the advantage of using Fermi-Bose transformation 
for solving Eq. \eqref{LLmodel}.

\section{The Lieb-Liniger gas in a wedgelike potential: Strongly interacting limit}
\label{sec:wedge}

In this section, we consider the Lieb-Liniger gas in the wedgelike potential 
defined as 
\begin{equation}
V(x) = \left\{ \begin{array}{ll}
         x & \mbox{if $x \geq 0$};\\
        \infty & \mbox{if $x < 0$}.\end{array} \right.
\label{wedge}
\end{equation}
For simplicity, we have fixed the value of the constant force to $\alpha=1$. 
Solutions for any other value can be obtained by simple rescaling: 
$x \rightarrow \alpha^{1/3}x$ and $E \rightarrow \alpha^{-2/3}E$. 

In order to find the ground state in such a potential, one should 
find solutions of Eqs. (\ref{LLmodel2}) and (\ref{interactions}) 
(assuming we work in the fundamental sector $R_1$), 
together with the following boundary condition: $\psi_{B,c}(x_1=0,x_2,\ldots,x_N)=0$. 
The first idea that may come to mind in attempting to find such a ground state 
is to utilize Eq. (\ref{ansatz}) as an ansatz, since it apparently 
obeys (\ref{LLmodel2}) and (\ref{interactions}), and try to adjust the $N$ free 
parameters $E_i$ such that $\psi_{B,c}(x_1=0,x_2,\ldots,x_N)=0$. 
Namely, such a procedure leads to the solutions for the ground states 
of a Lieb-Liniger gas on the ring \cite{Lieb1963}, where instead of 
the ansatz (\ref{ansatz}) with Airy functions, one utilizes an ansatz with 
plane waves, $\psi_{B,c}= {\mathcal N}_{c} \hat O_c \det_{m,j=1}^{N} e^{i k_j x_m}$ 
(e.g., see \cite{Korepin1993}), 
and instead of $E_j$, one adjusts the quasimomenta $k_j$ 
(which have to obey Bethe's equations) to acquire the proper boundary conditions. 
However, for this wedge like potential such a line of reasoning fails. 
Mathematically, this occurs because the first derivative of the Airy function 
is not simply related to the Airy function itself (whereas a derivative of a plane
wave is proportional to the plane wave itself). 

\begin{figure}
\centerline{
\mbox{\includegraphics[width=0.6\textwidth]{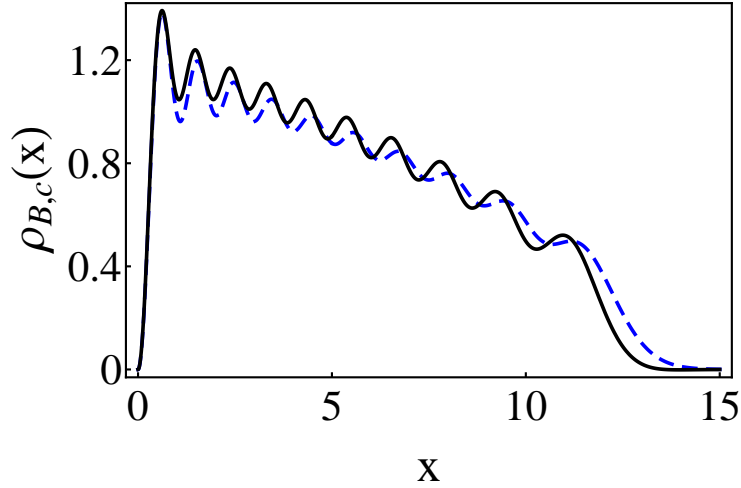}}
}
\caption{(Color online) The single particle density $\rho_{B,c}(x)$ 
(solid black line) of $N=10$ 
Lieb-Liniger bosons in a wedgelike potential ($c=40$, $\alpha=1$). 
Dashed blue line shows the density in the Tonks-Girardeau limit.
}
\label{fig:density}
\end{figure}

Nevertheless, we can find solutions in the form (\ref{ansatz}) in the 
Tonks-Girardeau limit ($c\rightarrow \infty$), and we can utilize some form of 
$1/c$ approximation to find deviations from the Tonks-Girardeau ground state 
for large but finite $c$. 
The Tonks-Girardeau ground state is constructed by symmetrizing the Slater determinant
of $N$ lowest single-particle eigenstates \cite{Girardeau1960}:
\begin{equation}
\psi_{TG} = \Pi_{k<m} \mbox{sgn}(x_m-x_k) 
\frac{1}{\sqrt{N!}} \det_{i,j=1}^{N} \phi_i(x_j),
\end{equation}
where 
\begin{equation}
\phi_i(x)=\frac{\mbox{Ai}(x-E_i)}{\mbox{Ai}^{\prime}(-E_i)}.
\end{equation}
The single-particle energies $E_i$ are such that $\mbox{Ai}(-E_i)=0$ [i.e., 
$\phi_i(0)=0$], and the eigenstates form an orthonormal set: 
$\int_0^\infty \phi_i^*(x)\phi_j(x)dx=\delta_{i,j}$. 
The ground-state energy is simply $E_{TG}=\sum_{i=1}^{N} E_i$.
As an illustration, in Fig. \ref{fig:density} we display the single-particle 
density for the Tonks-Girardeau ground state (dashed blue line) comprising 
$N=10$ particles. 

An approximative perturbative approach for calculating the properties of a 
Lieb-Liniger gas in the strongly interacting regime has been suggested by Sen 
\cite{Sen99-03}.
It can be shown that the perturbation around $c=\infty$ (the Tonks-Girardeau limit) 
is correctly described by a pseudopotential \cite{Sen99-03}
\begin{equation}
\hat{V}_{pp} = -\frac4c \sum_{i<j}\delta^{\prime \prime} (x_i - x_j),
\label{pp}
\end{equation}
that is, the pseudopotential (\ref{pp}) is utilized as a small perturbation 
around the Tonks-Girardeau ground state (unperturbed state) for large $c$.
It gives the correct first-order correction to the ground-state energy and wave 
function when plugged into the standard perturbation expressions with \(1/c\) 
as a small parameter. 

In the $1/c$ approximation, the ground-state energy of the Lieb-Liniger 
system is
\begin{align}
E_{B,c} &= E_{TG} +\left< \psi_{TG} \left| \hat{V}_{pp}  \right| 
\psi_{TG} \right> \nonumber \\
        &= E_{TG} - \frac1c N(N-1).
\label{energy_corr}
\end{align}
Result \eqref{energy_corr} is obtained by a direct calculation of the 
expectation value of the pseudopotential $\hat{V}_{pp}$ for the Tonks-Girardeau
ground state.
Such matrix elements are readily evaluated by using Slater-Condon rules:
\begin{align} 
\left< \psi_{TG} \left| \hat{V}_{pp}  \right| \psi_{TG} \right> =&
-\frac{4}{c} \sum_{i<j} \int _0^{\infty } dx 
\Big( 
\phi _i^*(x) \phi _i(x) \frac{d^2}{dy^2} [\phi_j^*(y)\phi _j(y)]_{y=x} \nonumber \\
 &-
\phi _i^*(x) \phi _j(x) \frac{d^2}{dy^2} [\phi _j^* (y) \phi _i(y)]_{y=x}
\Big).
\end{align}
We have verified \eqref{energy_corr} numerically (by employing Mathematica) 
up to $N = 20$ particles, and we conjecture that the expression is valid 
for any number of particles trapped by the potential \eqref{wedge}.

To first order in $1/c$, the Lieb-Liniger wave function is 
given by \cite{Sen99-03}
\begin{align} 
\psi_{B,c} \approx  \psi_{TG} 
       &+ \sum_{n \leq N,m>N} 
        \frac{ \left< \psi_{TG}^{(m;n)} \left| \hat{V}_{pp}  \right| 
	 \psi_{TG} \right>}{E_n-E_m}\, \psi_{TG}^{(m;n)} \nonumber \\
	&+ \sum_{ \substack{n<n^\prime \leq N \\ m^\prime>m>N}} 
	\frac{ \left< \psi_{TG}^{(m,m^\prime;n,n^\prime)} 
	                  \left| \hat{V}_{pp}  \right| 
	              \psi_{TG} \right>}{
	  E_n + E_{n^\prime}-E_m - E_{m^\prime}}
	\, \psi_{TG}^{(m,m^\prime;n,n^\prime)}, 	
\label{wf_corr}
\end{align}
where $\psi_{TG}^{(m;n)}$ labels an excited Tonks-Girardeau state; this state 
is obtained from the ground state $\psi_{TG}$ by replacing the single-particle 
state $\phi_n$, where $n \leq N$, with the single-particle state 
$\phi_m$ of higher energy, $m>N\geq n$. Analogously, 
\( \psi_{TG}^{(m,m^\prime;n,n^\prime)} \) labels two particle excitation of the
TG gas state.
The expression for the single-particle density 
$\rho_{B,c}(x)=N \int \! dx_2 \cdots dx_N |\psi_{B,c}|^2 $ 
can be calculated straightforwardly by employing the wave function from 
Eq. \eqref{wf_corr}, by keeping the terms up to $1/c$:  
\begin{align}
\rho_{B,c} (x) &\approx \rho_{TG} (x) + N
 \sum_{n \leq N,m>N} \left(\frac{ \left< \psi_{TG}^{(m;n)} 
                                     \left| \hat{V}_{pp} \right| 
                           \psi_{TG} \right>}{E_n-E_m}
\int \!\! dx_2 \cdots dx_N \, \psi^*_{TG} \psi_{TG}^{(m;n)} + 
c.c. \right)  \nonumber \\				
	 &\approx \rho_{TG} (x) + 
	   \frac1c \sum_{n \leq N,m>N} \frac{ V_{pp}^{(m;n)}}{E_n - E_m}\, 
	                      \phi_n (x) \phi_m (x).
\label{rhoLL}
\end{align}
Here, the matrix element
$V_{pp}^{(m;n)} \equiv -8 \left< \psi_{TG}^{(m;n)} \left| 
\sum_{i<j}\delta^{\prime \prime} (x_i - x_j) \right| \psi_{TG} \right>$
of the single-particle excitation 
from the level \(n \le N\) with energy \(E_n\), to the level \(m > N \) 
with energy \(E_m\) is given by
\begin{equation} 
V_{pp}^{(m;n)}=-8
\sum _{i=1,i\ne n}^N \int _0^{\infty } \!\! dx \Big( \phi _m^*(x) \phi _n(x)
\frac{d^2}{dy^2} [\phi _i^*(y)\phi _i(y)]_{y=x} 
 -\phi _m^*(x) \phi _i(x) 
\frac{d^2}{dy^2} [\phi _i^* (y) \phi _n(y)]_{y=x}\Big). 
\end{equation}
In Fig. \ref{fig:density} we illustrate the single-particle density 
$\rho_{B,c} (x)$ in $1/c$ approximation (solid black line), which is obtained 
by using Eq. (\ref{rhoLL}) for $N=10$ and $c=40$. It should be mentioned that 
the two-particle excitations [second sum in Eq. \eqref{wf_corr}] do not yield any 
contribution to the first-order single particle density $\rho_{B,c} (x)$, due 
to the vanishing of the overlap of the wave functions in calculation of the 
density (in the same way as demonstrated for the case of bosons confined in 
an infinitely deep box \cite{Sen99-03}). 
In our calculation of the density \( \rho_{B,c}\) via (\ref{rhoLL}), 
we have included only a finite number of terms, where the cutoff is chosen 
to be sufficiently large, such that the contribution of the remaining terms 
is negligible [for the calculation illustrated in 
Fig. \ref{fig:density}, we kept 150 terms in Eq. \eqref{rhoLL} with the
highest contribution].

\section{Exact quantum dynamics via a Fourier transform}
\label{sec:Ftran}
In this section we discuss the time-dependent solutions of the Lieb-Liniger system in a linear potential.
Before proceeding, we note that dynamics in the strongly interacting regime (i.e., 
dynamics of a Tonks-Girardeau gas in a linear potential) was studied in Ref. \cite{delCampo2006}.
Here, we assume that the bosons are initially localized by some external trapping potential. 
At time $t=0$, this potential is suddenly turned off, and bosons are released to 
evolve in the linear potential. This problem can be related to free expansion 
of the Lieb-Liniger wave packet by simple rescaling of the coordinates. 
If the wave function $\psi_{free}(x_1,\ldots,x_N,t)$ obeys the equation,
\begin{equation}
i \frac{\partial \psi_{free}}{\partial t}=
- \sum_{i=1}^{N}\frac{\partial^2 \psi_{free}}{\partial x_i^2} 
+ \sum_{1\leq i < j \leq N} 2c\,\delta(x_i-x_j)\psi_{free},
\label{free}
\end{equation}
(i.e., $\psi_{free}$ describes free expansion \cite{Jukic2008,Jukic2009}), 
then the wave function 
\begin{align}
\psi_{B,c}(x_1,\ldots,x_N,t) = e^{-i \alpha t 
\sum_{i=1}^{N} (x_i+\alpha t^2/3)} \ 
\psi_{free}(x_1+\alpha t^2,\ldots,x_N+\alpha t^2,t)
\label{translation}
\end{align}
is the solution of the time-dependent problem in the constant-force potential,
\begin{equation}
i \frac{\partial \psi_{B,c}}{\partial t}=
- \sum_{i=1}^{N}\frac{\partial^2 \psi_{B,c}}{\partial x_i^2} 
+ \sum_{1\leq i < j \leq N} 2c\,\delta(x_i-x_j)\psi_{B,c} 
+ \alpha \sum_{i=1}^N x_i  \psi_{B,c}.
\label{LLmodelLin}
\end{equation}
The initial conditions coincide (i.e., at $t=0$ we have $\psi_{B,c}=\psi_{free}$). 
Note that the phase factor in Eq. (\ref{translation})
accounts for the momentum per particle $\alpha t$, which is 
acquired in time in the field of constant force $\alpha$ (in units used here, $m=1/2$, 
and therefore the classical acceleration is $2 \alpha$).
Transformation (\ref{translation}) can be verified by direct substitution in Eq. (\ref{LLmodelLin}),
from which it becomes evident that it is valid for any two-particle interaction $V(x_i-x_j)$.
Namely, transformation $x_i \rightarrow x_i+\alpha t^2$ does not affect the two-particle 
interaction term $V(x_i-x_j)$ [in fact, because of this, Eq. (\ref{translation}) can be 
deduced from the well-known solution for a single-particle wave packet in a linear 
potential].

It is known that freely expanding Lieb-Liniger wave packets can be calculated 
by solving an $N$-dimensional Fourier transform \cite{Jukic2008}:
\begin{equation}
\psi_{free}(x_1,\ldots,x_N,t) = \int dk_1 \cdots dk_N 
G(k_1,\ldots,k_N) e^{i \sum_{i=1}^N \left( k_i x_i - k_i^2 t \right) }.
\label{freeSolution}
\end{equation}
We note that the function $G$ is \textit{not} the Fourier transform of
the wave function $\psi_{free}$
because it depends on the coordinates $x_j$ through the
$\mbox{sgn}(x_j-x_i)$ terms
(see Refs. \cite{Jukic2008,Jukic2009} for details), that is, it differs from one
permutation sector in $x$ space to the next. Nevertheless, by calculating
the integral
in Eq. (\ref{freeSolution}) in one sector (say $R_1$), we obtain
$\psi_{free}$ in
that sector, which is sufficient due to bosonic symmetry.
The function $G$ contains all information on initial conditions and it
can be expressed in terms of the projections of the initial wave function 
on the Lieb-Liniger free space eigenstates (e.g., see Refs. \cite{Jukic2008,Jukic2009}). 
By using Eqs. (\ref{freeSolution}) and (\ref{translation}) we can 
express $\psi_{B,c}$ in terms of an $N$-dimensional Fourier transform:
\begin{align}
\psi_{B,c}(x_1,\ldots,x_N,t) =& 
\int  dk_1 \cdots dk_N \nonumber \\
 & \times G(k_1,\ldots,k_N) 
\exp \left\{ i \sum_{i=1}^N \left[ (k_i-\alpha t) x_i + 
\frac{(k_i-\alpha t)^3-k_i^3}{3 \alpha} \right] \right\}.
\label{prefinal}
\end{align}

We would like to note that result (\ref{prefinal}) can be obtained by straightforward
use of Fermi-Bose transformation. The time-dependent wave function $\psi_F$ which describes
the system of $N$ noninteracting fermions in a linear potential $V(x)=\alpha x$ can be written
via its Airy transform:
\begin{equation}
\psi_F(x_1,\ldots,x_N,t)=\int dE_1 \cdots dE_N \bar{\psi}_F(E_1,\ldots,E_N)
e^{-it \sum_{i=1}^N E_i} \prod_{i=1}^N \mbox{Ai}(\alpha^{-2/3}(\alpha x_i - E_i)).
\label{solution}
\end{equation}
Here, $\bar{\psi}_F(E_1,\ldots,E_N)$ contains information on initial conditions,
\begin{equation}
\bar{\psi}_F(E_1,\ldots,E_N)=\left(\frac{1}{\alpha^{1/3}}\right)^N \int dx_1 \cdots dx_N  \psi_F(x_1,\ldots,x_N,0)
\prod_{i=1}^N \mbox{Ai}(\alpha^{-2/3}(\alpha x_i - E_i)).
\label{transform}
\end{equation}
By using the well-known relation between the Airy and Fourier transform $\tilde \psi_F$ \cite{Vallee},
\begin{equation}
\bar{\psi}_F(E_1,\ldots,E_N) =\left(\frac{1}{\alpha^{2/3}}\right)^N \int dk_1 \cdots dk_N \ 
\tilde{\psi}_F e^{i \sum_{i=1}^N \left(k_i E_i-k_i^3/3\right)/\alpha},
\end{equation}
we find
\begin{equation}
\psi_F(x_1,\ldots,x_N,t)= \int dk_1 \cdots dk_N \ \tilde{\psi}_F  \exp \left\{ i \sum_{i=1}^N 
\left[
(k_i-\alpha t)  x_i + \frac{(k_i-\alpha t)^3-k_i^3}{3 \alpha} \right] 
\right\}.
\label{psiFFourier}
\end{equation}
The time-dependent solution of the Lieb-Liniger model [i.e. Eq. \eqref{prefinal}], can now be found directly from
the expression above by applying the Fermi-Bose transformation operator $\hat O_c$ onto Eq. (\ref{psiFFourier}).

Our discussion in this section adds upon the previous studies of 
Lieb-Liniger wave-packet dynamics on an infinite line \cite{Gaudin1983,Buljan2008,Jukic2008,Jukic2009}, 
and in the presence of the hard-wall potential \cite{Jukic2010}; in all these cases 
the motion of an interacting Lieb-Liniger wave packet can be calculated by using an 
$N$-dimensional Fourier transform. 

In order to illustrate the connection between (\ref{free}) and (\ref{translation}), 
we present the following numerical example. 
The system of three Lieb-Liniger bosons are trapped in the ground state of an 
infinitely deep box of length $L=\pi$; at $t=0$, the trap is turned off 
and the bosons start to experience the constant force $\alpha=3$.
The exact initial wave function is constructed as a superposition of free space 
eigenstates \cite{Gaudin1971}. From this state we can find the 
function $G(k_1,\ldots,k_N)$ which keeps all information on initial conditions \cite{Jukic2008,Jukic2009}. 
By numerically calculating the integral in (\ref{prefinal}), we obtain the
time-dependent wave function $\psi_{B,c}(x_1,\ldots,x_N,t)$ describing the system. 
Here, we plot two relevant physical quantities,
the single-particle density, $\rho_{B,c}(x,t)=N \int dx_2 \cdots dx_N \left| \psi_{B,c}(x,\ldots,x_N,t)
\right|^2$, and the momentum distribution $n(k,t)$ (density in $k$ space).

\begin{figure}
\centerline{
\mbox{\includegraphics[width=0.35\textwidth]{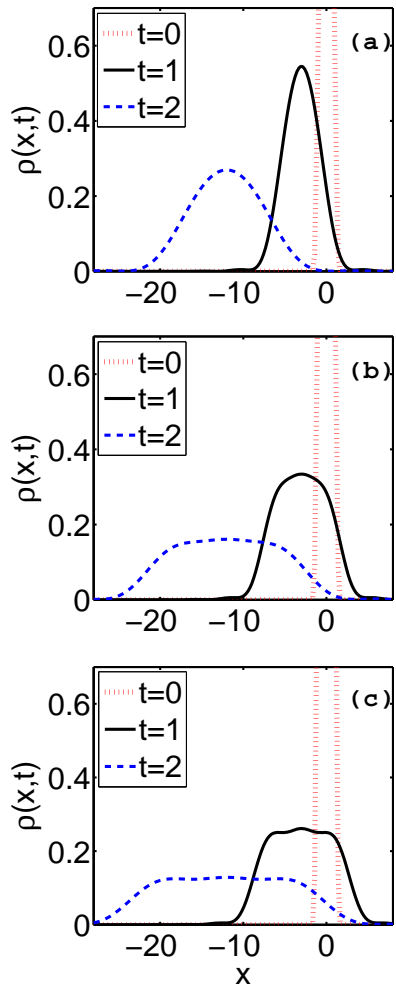}}
}
\caption{(Color online) Evolution of $N=3$ Lieb-Liniger bosons in the
linear potential $\alpha x$ ($\alpha=3$) from the ground state of a box
with infinitely high walls. Single-particle density in time for
various interaction strengths $c$: (a) $c=0.25$, (b) $c=3$, and (c) $c=10$.
Red dotted lines are for $t=0$, solid black lines are for $t=1$,
and blue dashed lines are for $t=2$.
}
\label{density}
\end{figure}

\begin{figure}
\centerline{
\mbox{\includegraphics[width=0.35\textwidth]{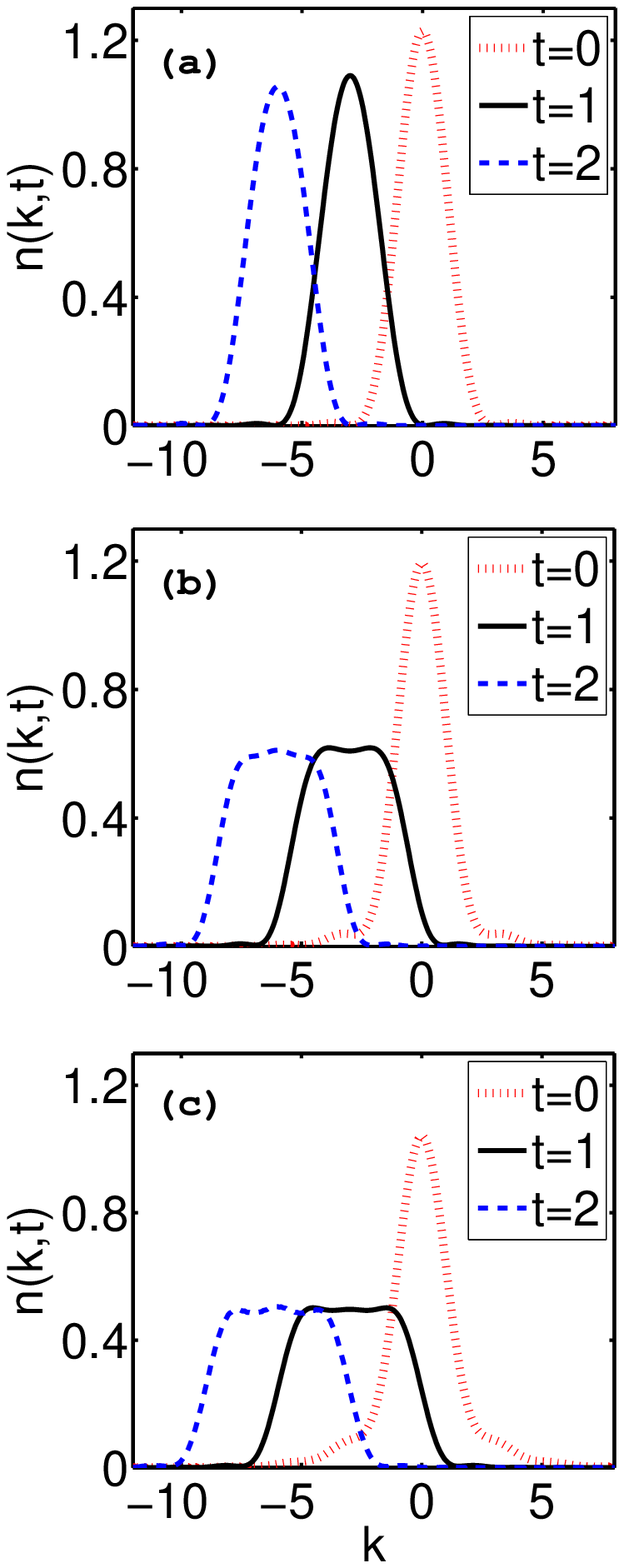}}
}
\caption{
(Color online) Evolution of the momentum distribution. The colors and lines for different
$c$ and $t$ are identical as in Fig. \ref{density}.  
}
\label{momentum}
\end{figure}

From Eq. (\ref{translation}) it follows that the density 
in coordinate space will be the same as in the case of free expansion ($\alpha=0$), with 
mere translation of the coordinates [$\rho_{B,c}(x,t)=\rho_{free}(x+\alpha t^2,t)$]. 
The momentum distribution will be equivalent also up to the simple 
transformation $k \rightarrow k-\alpha t$.
The density profile and momentum distribution
of the wave packet are plotted in Figs. \ref{density} and \ref{momentum}, respectively, for three various interactions strengths
$c$: (a) $c=0.25$, (b) $c=3$, and (c) $c=10$. Starting from $t=0$, the wave packet evolves
to the left in $x$ space with the center of mass motion $\alpha t^2$, while at the same time it spreads in width
independently. For large $c$, the spread is more pronounced, as can also be conjectured
from the initial momentum distribution. For very large $c$, the wave packet will asymptotically
experience fermionization of the momentum distribution \cite{Rigol2005,Minguzzi2005}.


%
%

\section{Conclusion}
\label{sec:concl}

We have studied the Lieb-Liniger model in the constant-force (linear) potential. 
Exact stationary solutions for this system, referred to as the  
Lieb-Liniger-Airy states, were constructed by employing Gaudin's Fermi-Bose mapping 
operator $\hat O_c$. This was enabled by the fact that the operator commutes 
with the linear potential: $[\hat O_c,\sum_j \alpha x_j]=0$. 
We have calculated the ground-state properties of the Lieb-Liniger gas, 
in the strongly interacting regime, in the wedgelike potential: 
$V(x)=\alpha x$ for $x>0$ ($\alpha>0$), and $V(x)=\infty$ for $x<0$. 
This was achieved in the Tonks-Girardeau regime and in $1/c$ approximation by 
employing the pseudopotential approach \cite{Sen99-03}.  
Finally, we have pointed out that the time-dependent Lieb-Liniger wave packets 
in the linear potential can be found by employing an $N$-dimensional 
Fourier transform.


\acknowledgments
The authors are grateful to Vladimir Gritsev for pointing out to them Ref. \cite{Sen1988}.
We acknowledge useful discussion with Thomas Gasenzer.  
This work is supported by the Croatian Ministry of Science (Grant No. 119-0000000-1015), 
the Croatian National Foundation for Science, and the Croatian-Israeli project cooperation. 


\end{document}